\documentstyle[preprint,aps]{revtex}
\tightenlines
\begin{document}
\draft
\title{On the dual topological quantum numbers filling factors}

\author{Wellington da Cruz}
\address{Departamento de F\'{\i}sica,\\
 Universidade Estadual de Londrina, Caixa Postal 6001,\\
Cep 86051-970 Londrina, PR, Brazil\\
E-mail address: wdacruz@exatas.uel.br}
\date{\today}
\maketitle
\begin{abstract}

We consider recent experimental results 
[W. Pan {\it et al}, Phys. Rev. Lett. {\bf 90}, 016801 (2003)] for 
occurrence of the fractional quantum Hall effect-FQHE under 
the perspective of our formulation in terms of {\it fractons}. These objects 
carry rational or irrational values of spin and satisfy a {\it fractal distribution 
function} associated with a {\it fractal von Neumann entropy}. 
According to our approach the 
{\it FQHE occurs in pairs of dual topological quantum numbers fillings factors} 
and this geometrical character comes from the {\it connection betwenn the fractal 
parameter or Hausdorff dimension $h$ and the spin $s$ of the particles}. We 
suggest to the experimentalists consider our 
ideas and verify in fact that this 
phenomenon of FQHE satisfies a {\it symmetry principle} discovered by us, i.e, 
{\it the duality symmetry betwenn universal classes of fractons}. 
\end{abstract}

\pacs{PACS numbers: 71.10.Pm, 05.30.-d, 73.43.Cd \\
Keywords:  Fractal distribution function; Fractal von Neumann entropy; 
Fractons; Fractional quantum Hall effect.}

\newpage

We have introduced in the literature\cite{R1,R2,R3,R4,R5,R6} the concept of universal classes $h$ 
of particles or quasiparticles, termed {\it fractons}, which satisfy 
a {\it fractal distribution function} associated with a {\it fractal 
von Newmann entropy}. These objects are charge-flux systems 
which carry rational or irrational 
values of spin. They are defined in 
two-dimensional multiply connected space and are classified in terms of the 
fractal parameter or Hausdorff dimension $h$ , defined in the interval 
$1$$\;$$ < $$\;$$h$$\;$$ <$$\;$$ 2$. The fractal properties of the 
quantum paths are characterized by $h$ and can be extracted from the propagators 
of the particles in the momentum space\cite{R1,R7}. Considering the {\it fractal spectrum}

\begin{eqnarray}
h-1=1-\nu,\;\;\;\; 0 < \nu < 1;\;\;\;\;\;\;\;\;
 h-1=\nu-1,\;
\;\;\;\;\;\; 1 <\nu < 2;\; etc.
\end{eqnarray}

\noindent and the {\it spin-statistics relation} $\nu=2s$, 
valid for such fractons, we can collect them in universal classes. 
{\it The fractal spectrum establishes a 
connection between $h$ and the spin $s$ of the particles}: 
$h=2-2s$, $0\leq s\leq \frac{1}{2}$. Thus, there exists a {\it mirror 
symmetry} behind this notion of fractal spectrum. For a review see\cite{R8}.

In a recent paper W. Pan {\it et al.}\cite{R9} present experimental 
results for occurrence of FQHE for some values of filling factors, 
$\nu=\frac{4}{11},
\frac{5}{13},\frac{7}{11},\frac{4}{13},\frac{6}{17},\frac{5}{17}$. Thus, 
in this Letter we 
make some comments about these values under the perspective of 
our approach to the FQHE in terms of fractons. 
According to our formulation the {\it FQHE occurs in pairs of dual 
topological 
quantum numbers filling factors}. These ones are numerically equal to statistical 
parameter $\nu$. So, {\it the topological 
character of these 
quantum numbers filling factors comes from the connection with the Hausdorff dimension 
$h$}.

We have found that the filling factors related to the 
quantization of the Hall resistance in systems of the condensed matter 
under extreme conditions of 
temperature and magnetic field\cite{R10} can be classified in terms 
of the Hausdorff dimension associated with the quantum paths\cite{R1}. This is so, 
because {\it there exists a symmetry principle, i.e, the duality symmetry 
betwenn universal 
classes of fractons, defined by ${\tilde{h}}=3-h$}. As a consequence, 
we extract a {\it fractal supersymmetry} which defines pairs of particles 
$\left(s,s+\frac{1}{2}\right)$. 

This way, we discovered that the quantum Hall transitions satisfy some 
properties related with the Farey sequences of rational numbers and the 
transitions allowed are those generated by the condition
 $\mid p_{2}q_{1}
-p_{1}q_{2}\mid=1$, 
with $\nu_{1}=\frac{p_{1}}{q_{1}}$ and $\nu_{2}=
\frac{p_{2}}{q_{2}}$.

For example, we have the {\it possible transitions} between the classes

\[
\biggl\{\frac{1}{3},\frac{5}{3},\frac{7}{3},\cdots\biggr\}_{h=\frac{5}{3}}\rightarrow
\biggl\{\frac{2}{5},\frac{8}{5},\frac{12}{5},\cdots\biggr\}_{h=\frac{8}{5}}\rightarrow
\biggl\{\frac{3}{7},\frac{11}{7},\frac{17}{7},\cdots\biggr\}_{h=\frac{11}{7}}\rightarrow\]
\[
\biggl\{\frac{4}{9},\frac{14}{9},\frac{22}{9},\cdots\biggr\}_{h=\frac{14}{9}}\rightarrow
\biggl\{\frac{5}{11},\frac{17}{11},\frac{27}{11},\cdots\biggr\}_{h=\frac{17}{11}}
\rightarrow
\biggl\{\frac{6}{13},\frac{20}{13},\frac{32}{13},\cdots\biggr\}_{h=\frac{20}{13}} etc.
\]

\noindent and for their respective duals 

\[
\biggl\{\frac{2}{3},\frac{4}{3},\frac{8}{3},\cdots\biggr\}_{h=\frac{4}{3}}\rightarrow
\biggl\{\frac{3}{5},\frac{7}{5},\frac{13}{5},\cdots\biggr\}_{h=\frac{7}{5}}\rightarrow
\biggl\{\frac{4}{7},\frac{10}{7},\frac{18}{7},\cdots\biggr\}_{h=\frac{10}{7}}\rightarrow\]
\[
\biggl\{\frac{5}{9},\frac{13}{9},\frac{23}{9},\cdots\biggr\}_{h=\frac{13}{9}}\rightarrow
\biggl\{\frac{6}{11},\frac{16}{11},\frac{28}{11},\cdots\biggr\}_{h=\frac{16}{11}}
\rightarrow
\biggl\{\frac{7}{13},\frac{19}{13},\frac{33}{13},\cdots\biggr\}_{h=\frac{19}{13}} etc.
\]

\noindent Thus, we obtain {\it the universality classes of the quantum Hall 
transitions}. Now, follows some examples of these pairs of {\it dual topological quantum numbers} 
just observed

\begin{eqnarray} 
\label{e.22}
(\nu,\tilde{\nu})&=&\left(\frac{1}{3},\frac{2}{3}\right), 
\left(\frac{5}{3},\frac{4}{3}\right), \left(\frac{1}{5},\frac{4}{5}\right), 
\left(\frac{2}{7},\frac{5}{7}\right),\left(\frac{2}{9},\frac{7}{9}\right), 
\left(\frac{2}{5},\frac{3}{5}\right),\left(\frac{7}{3},\frac{8}{3}\right),\nonumber\\
&&\left(\frac{3}{7},\frac{4}{7}\right), 
\left(\frac{4}{9},\frac{5}{9}\right), \left(\frac{8}{5},\frac{7}{5}\right),
\left(\frac{6}{13},\frac{7}{13}\right),\left(\frac{5}{11},\frac{6}{11}\right),
\left(\frac{10}{7},\frac{11}{7}\right),\\
&&\left(\frac{4}{11},\frac{7}{11}\right),
\left(\frac{13}{4},\frac{15}{4}\right), \left(\frac{14}{5},\frac{11}{5}\right),
\left(\frac{9}{4},\frac{11}{4}\right),\left(\frac{19}{7},\frac{16}{7}\right) 
etc.\nonumber
\end{eqnarray}

\noindent We claim that {\bf all the experimental results satisfy 
this symmetry principle}. In this way, {\it our formulatiom can 
predicting FQHE, that is, consider the duality 
symmetry discovered by us}\cite{R1}. We also observe that 
{\it our approach, in terms of equivalence 
classes for the filling factors, embodies 
the structure of the modular group} as discussed in the literature
\cite{R1,R11}.

In this note we suggest to the experimentalists\cite{R9} consider our 
ideas and verify in fact that the FQHE occurs for the duals 
$\left(\frac{5}{13},\frac{8}{13}\right),\left(\frac{4}{13},\frac{9}{13}\right),
\left(\frac{6}{17},\frac{11}{17}\right),\left(\frac{5}{17},\frac{12}{17}\right)$. 
They have found in\cite{R9}, the dual pair 
$\left(\frac{4}{11},\frac{7}{11}\right)$, which confirms again our 
symmetry principle.
 
According to the rules discussed above for the quantum Hall 
transitions, the results of the Ref.\cite{R9} can be written in the form:

\begin{eqnarray}
\label{e.3}
&&\frac{5}{7}\rightarrow\frac{12}{17}\rightarrow\frac{7}{10}\rightarrow
\frac{9}{13}\rightarrow\frac{2}{3}\rightarrow\frac{11}{17}\rightarrow\frac{9}{14}
\rightarrow\frac{7}{11}\rightarrow\frac{5}{8}\rightarrow\frac{8}{13}
\rightarrow\frac{3}{5}\rightarrow\nonumber\\
&&\frac{4}{7}\rightarrow\frac{5}{9}\rightarrow\frac{6}{11}
\rightarrow\frac{7}{13}\rightarrow\frac{8}{15}\rightarrow\frac{9}{17}
\rightarrow\frac{10}{19}\rightarrow\frac{11}{21}\rightarrow\frac{1}{2}
\rightarrow\frac{10}{21}\rightarrow\frac{9}{19}\rightarrow\nonumber\\
&&\frac{8}{17}\rightarrow\frac{7}{15}\rightarrow\frac{6}{13}
\rightarrow\frac{5}{11}\rightarrow\frac{4}{9}\rightarrow\frac{3}{7}
\rightarrow\frac{2}{5}
\rightarrow\frac{5}{13}\rightarrow\frac{3}{8}\rightarrow\frac{4}{11}
\rightarrow\frac{5}{14}\rightarrow\\
&&\frac{6}{17}\rightarrow\frac{1}{3}\rightarrow\frac{4}{13}
\rightarrow\frac{3}{10}\rightarrow\frac{5}{17}\rightarrow\frac{2}{7}.\nonumber 
\end{eqnarray}

\noindent As we can see, other sequences follow from this one, 
just take into account the rules. Now, considering the fractal 
spectrum and the duality symmetry 
betwenn the classes, we can determine the universality 
classes associated to this sequence Eq.(\ref{e.3}):

\begin{eqnarray}
&&\biggl\{\frac{5}{7},\frac{9}{7},\frac{19}{7},\cdots\biggr\}_{h=\frac{9}{7}}
\rightarrow\;\;\;\;
\biggl\{\frac{12}{17},\frac{22}{17},\frac{46}{17},
\cdots\biggr\}_{h=\frac{22}{17}}\rightarrow
\biggl\{\frac{7}{10},\frac{13}{10},\frac{27}{10},\cdots\biggr\}_{h=\frac{13}{10}}
\rightarrow\nonumber\\
&&\biggl\{\frac{9}{13},\frac{17}{13},\frac{35}{13},\cdots\biggr\}_{h=\frac{17}{13}}
\rightarrow
\biggl\{\frac{2}{3},\frac{4}{3},\frac{8}{3},\cdots\biggr\}_{h=\frac{4}{3}}
\rightarrow\;\;\;\;\;\;
\biggl\{\frac{11}{17},\frac{23}{17},\frac{45}{17},\cdots\biggr\}_{h=\frac{23}{17}}
\rightarrow\nonumber\\
&&\biggl\{\frac{9}{14},\frac{19}{14},\frac{37}{14},\cdots\biggr\}_{h=\frac{19}{14}}
\rightarrow
\biggl\{\frac{7}{11},\frac{15}{11},\frac{29}{11},\cdots\biggr\}_{h=\frac{15}{11}}
\rightarrow\biggl\{\frac{5}{8},\frac{11}{8},\frac{21}{8},\cdots\biggr\}_{h=\frac{11}{8}}
\rightarrow\nonumber\\
&&\biggl\{\frac{8}{13},\frac{18}{13},\frac{34}{13},\cdots\biggr\}_{h=\frac{18}{13}}
\rightarrow
\biggl\{\frac{3}{5},\frac{7}{5},\frac{13}{5},\cdots\biggr\}_{h=\frac{7}{5}}
\rightarrow\;\;\;\;\;
\biggl\{\frac{4}{7},\frac{10}{7},\frac{18}{7},\cdots\biggr\}_{h=\frac{10}{7}}
\rightarrow\nonumber\\
&&\biggl\{\frac{5}{9},\frac{13}{9},\frac{23}{9},\cdots\biggr\}_{h=\frac{13}{9}}
\rightarrow\;\;
\biggl\{\frac{6}{11},\frac{16}{11},\frac{28}{11},\cdots\biggr\}_{h=\frac{16}{11}}
\rightarrow
\biggl\{\frac{7}{13},\frac{19}{13},\frac{33}{13},\cdots\biggr\}_{h=\frac{19}{13}}
\rightarrow\nonumber\\
&&\biggl\{\frac{8}{15},\frac{22}{15},\frac{38}{15},\cdots\biggr\}_{h=\frac{22}{15}}
\rightarrow
\biggl\{\frac{9}{17},\frac{25}{17},\frac{43}{17},\cdots\biggr\}_{h=\frac{25}{17}}
\rightarrow
\biggl\{\frac{10}{19},\frac{28}{19},\frac{48}{19},\cdots\biggr\}_{h=\frac{28}{19}}
\rightarrow\\
&&\biggl\{\frac{11}{21},\frac{31}{21},\frac{53}{21},\cdots\biggr\}_{h=\frac{31}{21}}
\rightarrow
\biggl\{\frac{1}{2},\frac{3}{2},\frac{5}{2},\cdots\biggr\}_{h=\frac{3}{2}}
\rightarrow\;\;\;\;\;\;
\biggl\{\frac{10}{21},\frac{32}{21},\frac{52}{21},\cdots\biggr\}_{h=\frac{32}{21}}
\rightarrow\nonumber\\
&&\biggl\{\frac{9}{19},\frac{29}{19},\frac{47}{19},\cdots\biggr\}_{h=\frac{29}{19}}
\rightarrow
\biggl\{\frac{8}{17},\frac{26}{17},\frac{42}{17},\cdots\biggr\}_{h=\frac{26}{17}}
\rightarrow
\biggl\{\frac{7}{15},\frac{23}{15},\frac{37}{15},\cdots\biggr\}_{h=\frac{23}{15}}
\rightarrow\nonumber\\
&&\biggl\{\frac{6}{13},\frac{20}{13},\frac{32}{13},\cdots\biggr\}_{h=\frac{20}{13}}
\rightarrow
\biggl\{\frac{5}{11},\frac{17}{11},\frac{27}{11},\cdots\biggr\}_{h=\frac{17}{11}}
\rightarrow
\biggl\{\frac{4}{9},\frac{14}{9},\frac{22}{9},\cdots\biggr\}_{h=\frac{14}{9}}
\rightarrow\nonumber\\
&&\biggl\{\frac{3}{7},\frac{11}{7},\frac{17}{7},\cdots\biggr\}_{h=\frac{11}{7}}
\rightarrow\;\;
\biggl\{\frac{2}{5},\frac{8}{5},\frac{12}{5},\cdots\biggr\}_{h=\frac{8}{5}}
\rightarrow\;\;\;\;
\biggl\{\frac{5}{13},\frac{21}{13},\frac{31}{13},\cdots\biggr\}_{h=\frac{21}{13}}
\rightarrow\nonumber\\
&&\biggl\{\frac{3}{8},\frac{13}{8},\frac{19}{8},\cdots\biggr\}_{h=\frac{13}{8}}
\rightarrow\;\;
\biggl\{\frac{4}{11},\frac{18}{11},\frac{26}{11},\cdots\biggr\}_{h=\frac{18}{11}}
\rightarrow
\biggl\{\frac{5}{14},\frac{23}{14},\frac{33}{14},\cdots\biggr\}_{h=\frac{23}{14}}
\rightarrow\nonumber\\
&&\biggl\{\frac{6}{17},\frac{28}{17},\frac{40}{17},\cdots\biggr\}_{h=\frac{28}{17}}
\rightarrow
\biggl\{\frac{1}{3},\frac{5}{3},\frac{7}{3},\cdots\biggr\}_{h=\frac{5}{3}}
\rightarrow\;\;\;\;\;\;\;
\biggl\{\frac{4}{13},\frac{22}{13},\frac{30}{13},\cdots\biggr\}_{h=\frac{22}{13}}
\rightarrow\nonumber\\
&&\biggl\{\frac{3}{10},\frac{17}{10},\frac{23}{10},\cdots\biggr\}_{h=\frac{17}{10}}
\rightarrow
\biggl\{\frac{5}{17},\frac{29}{17},\frac{39}{17},\cdots\biggr\}_{h=\frac{29}{17}}
\rightarrow
\biggl\{\frac{2}{7},\frac{12}{7},\frac{16}{7},\cdots\biggr\}_{h=\frac{12}{7}}.
\nonumber
\end{eqnarray}

\noindent We observe that the duals of $\nu=\frac{5}{13},\frac{6}{17},
\frac{4}{13},\frac{5}{17}$, i.e, ${\tilde{\nu}}=
\frac{8}{13},\frac{11}{17},\frac{9}{13},\frac{12}{17}$, 
appear in the sequence obtained by our approach and for these values of 
filling factors we can verify the occurrence of FQHE.

Finally, as we can check, the formulation advocates by us, 
contrary other one\cite{R12}, offers {\it a rationale} for all observed 
filling factors. Again, we emphasize that 
{\it ours objects are spinning particles} and we believe that {\it this 
characteristic is crucial for understand the Hall states associated with fractons}. 
On the other hand, in our formulation, 
the filling factors are classified in classes labeled by the Hausdorff dimension 
associated with the spin of the particles. {\it The quantum Hall 
transitions satisfy some rules supported by symmetry principles, 
such as duality symmetry betwenn the universal classes of fractons and 
the transitions allowed are those that satisfy a mathematical structure which comes 
from the modular group}\cite{R1,R11}.

Once more, the FQHE occurs 
in pairs of dual topological quantum numbers filling factors and 
our approach can predicting such phenomena: just consider any value 
observed, {\it take into account the duality symmetry and the fractal 
spectrum}, and we receive the respective dual. All the experimental 
results obey these rules, i.e, {\it symmetry principles which 
support the theoretical basis of our formulation}. {\it The sequences 
observed for occurrence of FQHE appear naturally from first principles}, 
i.e, our approach does not has an empirical character. {\it It has a fundamental meaning 
posed by the connection betwenn a geometrical parameter $h$ associated with the 
quantum paths and the spin $s$ of the particles}. Then, this constitutes {\bf the 
great difference} betwenn our insight and the others of the literature. 
On the other hand, we do not have a Hamiltonian for the FQHE, however {\it we have 
symmetries associated with it and this can gives us 
some hints to solve this task}. Hence, this sounds for us a convenient strategy. Also, 
in any case, the system of particles which we are considering 
to characterize the FQHE satisfies a fractal distribution function 
associated with a fractal Von Neumann entropy and this way 
the physical scenario takes place ( see \cite{R8} for crucial details ). Besides this, all our theoretical description is 
supported by symmetry principles as we emphasized a lot along of our argumentation. 
The result of our formulation is very clear, we have established 
a fractal-like structure for the FQHE in terms of concepts of the 
fractal geometry as we discussed in\cite{R17} and 
so, some comments in\cite{R18} about the fractal 
approach to the FQHE was antecipated by us in the just 
way\cite{R1,R2,R3,R4,R5,R6,R8}.

As a final example, consider the results of the Refs.\cite{R9,R13,R14,R15,R16} 
together, so we can write the sequence: 

\begin{eqnarray}
&&\frac{15}{4}\rightarrow\frac{11}{3}\rightarrow\;
\frac{7}{2}\rightarrow\;\frac{10}{3}\rightarrow
\frac{13}{4}\rightarrow\;{\bf \frac{3}{1}}\rightarrow\;
\frac{14}{5}\rightarrow\frac{11}{4}\rightarrow
\frac{19}{7}\rightarrow\;\frac{8}{3}\rightarrow
\;\frac{13}{5}\rightarrow\;\frac{18}{7}\rightarrow\nonumber\\
&&\frac{23}{9}\rightarrow\frac{28}{11}\rightarrow
\frac{33}{13}\rightarrow\;\frac{5}{2}\rightarrow
\;\frac{32}{13}\rightarrow\frac{27}{11}\rightarrow
\;\frac{22}{9}\rightarrow\frac{17}{7}\rightarrow
\frac{12}{5}\rightarrow\;\frac{7}{3}\rightarrow
\;\frac{16}{7}\rightarrow\;\;\frac{9}{4}\rightarrow\nonumber\\
&&\frac{11}{5}\rightarrow{\bf \frac{2}{1}}\rightarrow
\;\frac{15}{8}\rightarrow\;\frac{28}{15}\rightarrow
\frac{13}{7}\rightarrow\frac{11}{6}\rightarrow\;
\frac{9}{5}\rightarrow\;\frac{16}{9}\rightarrow\;
\frac{7}{4}\rightarrow\;\frac{19}{11}\rightarrow
\;\frac{12}{7}\rightarrow\;\frac{17}{10}\rightarrow\nonumber\\
&&\frac{5}{3}\rightarrow\;\;\frac{8}{5}\rightarrow\;
\frac{11}{7}\rightarrow\;\frac{14}{9}\rightarrow
\frac{17}{11}\rightarrow\frac{20}{13}\rightarrow\;
\frac{3}{2}\rightarrow\;\frac{19}{13}\rightarrow
\frac{16}{11}\rightarrow\frac{13}{9}\rightarrow
\frac{10}{7}\rightarrow\;\;\frac{7}{5}\rightarrow\nonumber\\
&&\frac{4}{3}\;\rightarrow\frac{13}{10}\rightarrow\;
\frac{9}{7}\rightarrow\;\;\frac{14}{11}\rightarrow\;
\frac{5}{4}\rightarrow\;\frac{11}{9}\rightarrow\;
\frac{6}{5}\rightarrow\;\;\frac{7}{6}\rightarrow\;
\;\frac{8}{7}\rightarrow\;\frac{17}{15}\rightarrow
\frac{26}{23}\rightarrow\;\;\frac{9}{8}\rightarrow\nonumber\\
&&{\bf\frac{1}{1}}\rightarrow\;\;\frac{8}{9}\;\rightarrow
\frac{15}{17}\rightarrow\;\frac{7}{8}\rightarrow
\;\;\frac{13}{15}\rightarrow\;\frac{6}{7}\rightarrow\;
\frac{11}{13}\rightarrow\frac{16}{19}\rightarrow\;
\frac{5}{6}\rightarrow\;\frac{19}{23}\rightarrow
\frac{14}{17}\rightarrow\;\frac{9}{11}\rightarrow\\
&&\frac{4}{5}\rightarrow\;\;\frac{7}{9}\;\rightarrow
\frac{10}{13}\rightarrow\;\frac{13}{17}\rightarrow
\frac{16}{21}\rightarrow\;\frac{19}{25}\rightarrow\;
\frac{3}{4}\rightarrow\;\frac{14}{19}\rightarrow
\;\frac{11}{15}\rightarrow\frac{8}{11}\rightarrow
\frac{5}{7}\rightarrow\;\frac{12}{17}\rightarrow\nonumber\\
&&\frac{7}{10}\rightarrow\frac{9}{13}\rightarrow
\frac{2}{3}\rightarrow\;\;\frac{11}{17}\rightarrow
\frac{9}{14}\rightarrow\;\frac{7}{11}\rightarrow\;
\frac{5}{8}\rightarrow\;\frac{8}{13}\rightarrow
\;\;\frac{3}{5}\rightarrow\;\;\frac{4}{7}\rightarrow
\;\frac{5}{9}\rightarrow\;\frac{6}{11}\rightarrow\nonumber\\
&&\frac{7}{13}\rightarrow\frac{8}{15}\rightarrow
\frac{9}{17}\rightarrow\frac{10}{19}\rightarrow\;
\frac{11}{21}\rightarrow\;\;\frac{1}{2}\rightarrow\;
\frac{10}{21}\rightarrow\frac{9}{19}\rightarrow
\;\frac{8}{17}\rightarrow\frac{7}{15}\rightarrow
\frac{6}{13}\rightarrow\frac{5}{11}\rightarrow\nonumber\\
&&\frac{4}{9}\rightarrow\;\;\frac{3}{7}\rightarrow
\;\;\;\frac{2}{5}\rightarrow\;\frac{5}{13}\rightarrow
\;\frac{3}{8}\rightarrow\;\;\frac{4}{11}\rightarrow
\frac{5}{14}\rightarrow\frac{6}{17}\rightarrow\;
\;\frac{1}{3}\rightarrow\;\frac{4}{13}\rightarrow
\frac{3}{10}\rightarrow\frac{5}{17}\rightarrow\nonumber\\
&&\frac{2}{7}\rightarrow\;\frac{3}{11}\rightarrow\;
\frac{4}{15}\rightarrow\;\frac{5}{19}\rightarrow
\frac{6}{23}\rightarrow\;\;\frac{1}{4}\rightarrow\;
\frac{6}{25}\rightarrow\frac{5}{21}\rightarrow\;
\frac{4}{17}\rightarrow\frac{3}{13}\rightarrow
\frac{2}{9}\rightarrow\;\;\frac{1}{5}\rightarrow\nonumber\\
&&\frac{2}{11}\rightarrow\frac{3}{17}\rightarrow\;
\frac{4}{23}\rightarrow\;\frac{1}{6}\rightarrow
\;\frac{3}{19}\rightarrow\;\frac{2}{13}\rightarrow\;
\frac{1}{7}\rightarrow\;\frac{2}{15}\rightarrow
\;\;\frac{1}{8}\rightarrow\;\frac{2}{17}\rightarrow
\frac{1}{9}.\nonumber 
\end{eqnarray}

\noindent Thus, we identify dual pairs of 
filling factors observed ( see Eq.(\ref{e.22}) ) and other ones to be verified, as 
$(\nu,{\tilde{\nu}})=(\frac{2}{11},\frac{9}{11}), (\frac{3}{19},\frac{16}{19}), 
(\frac{3}{17},\frac{14}{17}), (\frac{2}{13},\frac{11}{13}), 
(\frac{1}{7},\frac{6}{7}), (\frac{2}{15},\frac{13}{15}), 
(\frac{2}{17},\frac{15}{17}), (\frac{1}{9},\frac{8}{9})\; etc. $

Another interesting point about our discussion is that for the sequence Eq.(5) we 
verify distinct possible transitions for the 
fractional quantum Hall effect, we have family or 
group of universality classes, for example, consider the {\bf Group IV}:

\newpage

\begin{eqnarray}
&&\biggl\{\frac{8}{9},\frac{10}{9},\frac{26}{9},\frac{28}{9},
\cdots\biggr\}_{h=\frac{10}{9}}
\rightarrow\;\;
\biggl\{\frac{15}{17},\frac{19}{17},\frac{49}{17},\frac{53}{17},
\cdots\biggr\}_{h=\frac{19}{17}}
\rightarrow\biggl\{\frac{7}{8},\frac{9}{8},\frac{23}{8},\frac{25}{8},
\cdots\biggr\}_{h=\frac{9}{8}}\rightarrow\nonumber\\
&&\biggl\{\frac{13}{15},\frac{17}{15},\frac{43}{15},\frac{47}{15},
\cdots\biggr\}_{h=\frac{17}{15}}
\rightarrow
\biggl\{\frac{6}{7},\frac{8}{7},\frac{20}{7},\frac{22}{7},
\cdots\biggr\}_{h=\frac{8}{7}}
\rightarrow\;\;\;\;\;\biggl\{\frac{11}{13},\frac{15}{13},\frac{37}{13},\frac{41}{13},
\cdots\biggr\}_{h=\frac{15}{13}}\rightarrow\nonumber\\
&&\biggl\{\frac{16}{19},\frac{22}{19},\frac{54}{19},\frac{60}{19},
\cdots\biggr\}_{h=\frac{22}{19}}
\rightarrow
\biggl\{\frac{5}{6},\frac{7}{6},\frac{17}{6},\frac{19}{6},
\cdots\biggr\}_{h=\frac{7}{6}}
\rightarrow\;\;\;\;\;\biggl\{\frac{19}{23},\frac{27}{23},\frac{65}{23},\frac{73}{23},
\cdots\biggr\}_{h=\frac{27}{23}}\rightarrow\nonumber\\
&&\biggl\{\frac{14}{17},\frac{20}{17},\frac{48}{17},\frac{54}{17},
\cdots\biggr\}_{h=\frac{20}{17}}
\rightarrow
\biggl\{\frac{9}{11},\frac{13}{11},\frac{31}{11},\frac{35}{11},
\cdots\biggr\}_{h=\frac{13}{11}}
\rightarrow\biggl\{\frac{4}{5},\frac{6}{5},\frac{14}{5},\frac{16}{5},
\cdots\biggr\}_{h=\frac{6}{5}}\rightarrow\nonumber\\
&&\biggl\{\frac{7}{9},\frac{11}{9},\frac{25}{9},\frac{29}{9},
\cdots\biggr\}_{h=\frac{11}{9}}
\rightarrow\;\;
\biggl\{\frac{10}{13},\frac{16}{13},\frac{36}{13},\frac{42}{13},
\cdots\biggr\}_{h=\frac{16}{13}}
\rightarrow\biggl\{\frac{13}{17},\frac{21}{17},\frac{47}{17},\frac{55}{17},
\cdots\biggr\}_{h=\frac{21}{17}}\rightarrow\nonumber\\
&&\biggl\{\frac{16}{21},\frac{26}{21},\frac{58}{21},\frac{68}{21},
\cdots\biggr\}_{h=\frac{26}{21}}
\rightarrow
\biggl\{\frac{19}{25},\frac{31}{25},\frac{69}{25},\frac{81}{25},
\cdots\biggr\}_{h=\frac{31}{25}}
\rightarrow\biggl\{\frac{3}{4},\frac{5}{4},\frac{11}{4},\frac{13}{4},
\cdots\biggr\}_{h=\frac{5}{4}}\rightarrow\nonumber\\
&&\biggl\{\frac{14}{19},\frac{24}{19},\frac{52}{19},\frac{62}{19},
\cdots\biggr\}_{h=\frac{24}{19}}
\rightarrow
\biggl\{\frac{11}{15},\frac{19}{15},\frac{41}{15},\frac{49}{15},
\cdots\biggr\}_{h=\frac{19}{15}}
\rightarrow\biggl\{\frac{8}{11},\frac{14}{11},\frac{30}{11},\frac{36}{11},
\cdots\biggr\}_{h=\frac{14}{11}}\rightarrow\nonumber\\
&&\biggl\{\frac{5}{7},\frac{9}{7},\frac{19}{7},\frac{23}{7},
\cdots\biggr\}_{h=\frac{9}{7}}\rightarrow\;\;\;\;\;
\biggl\{\frac{12}{17},\frac{22}{17},\frac{46}{17},\frac{56}{17},
\cdots\biggr\}_{h=\frac{22}{17}}\rightarrow
\biggl\{\frac{7}{10},\frac{13}{10},\frac{27}{10},\frac{33}{10},
\cdots\biggr\}_{h=\frac{13}{10}}
\rightarrow\nonumber\\
&&\biggl\{\frac{9}{13},\frac{17}{13},\frac{35}{13},\frac{43}{13},
\cdots\biggr\}_{h=\frac{17}{13}}
\rightarrow
\biggl\{\frac{2}{3},\frac{4}{3},\frac{8}{3},\frac{10}{3},
\cdots\biggr\}_{h=\frac{4}{3}}
\rightarrow\;\;\;\;\;\;\;
\biggl\{\frac{11}{17},\frac{23}{17},\frac{45}{17},\frac{57}{17},
\cdots\biggr\}_{h=\frac{23}{17}}
\rightarrow\nonumber\\
&&\biggl\{\frac{9}{14},\frac{19}{14},\frac{37}{14},\frac{47}{14},
\cdots\biggr\}_{h=\frac{19}{14}}
\rightarrow
\biggl\{\frac{7}{11},\frac{15}{11},\frac{29}{11},\frac{37}{11},
\cdots\biggr\}_{h=\frac{15}{11}}
\rightarrow\biggl\{\frac{5}{8},\frac{11}{8},\frac{21}{8},\frac{27}{8},
\cdots\biggr\}_{h=\frac{11}{8}}
\rightarrow\nonumber\\
&&\biggl\{\frac{8}{13},\frac{18}{13},\frac{34}{13},\frac{44}{13},
\cdots\biggr\}_{h=\frac{18}{13}}
\rightarrow
\biggl\{\frac{3}{5},\frac{7}{5},\frac{13}{5},\frac{17}{5},
\cdots\biggr\}_{h=\frac{7}{5}}
\rightarrow\;\;\;\;\;
\biggl\{\frac{4}{7},\frac{10}{7},\frac{18}{7},\frac{24}{7},
\cdots\biggr\}_{h=\frac{10}{7}}
\rightarrow\nonumber\\
&&\biggl\{\frac{5}{9},\frac{13}{9},\frac{23}{9},\frac{31}{9},
\cdots\biggr\}_{h=\frac{13}{9}}
\rightarrow\;\;
\biggl\{\frac{6}{11},\frac{16}{11},\frac{28}{11},\frac{38}{11},
\cdots\biggr\}_{h=\frac{16}{11}}
\rightarrow
\biggl\{\frac{7}{13},\frac{19}{13},\frac{33}{13},\frac{45}{13},
\cdots\biggr\}_{h=\frac{19}{13}}
\rightarrow\nonumber\\
&&\biggl\{\frac{8}{15},\frac{22}{15},\frac{38}{15},\frac{52}{15},
\cdots\biggr\}_{h=\frac{22}{15}}
\rightarrow
\biggl\{\frac{9}{17},\frac{25}{17},\frac{43}{17},\frac{59}{17},
\cdots\biggr\}_{h=\frac{25}{17}}
\rightarrow
\biggl\{\frac{10}{19},\frac{28}{19},\frac{48}{19},\frac{66}{19},
\cdots\biggr\}_{h=\frac{28}{19}}
\rightarrow\\
&&\biggl\{\frac{11}{21},\frac{31}{21},\frac{53}{21},\frac{73}{21},
\cdots\biggr\}_{h=\frac{31}{21}}
\rightarrow
\biggl\{\frac{1}{2},\frac{3}{2},\frac{5}{2},\frac{7}{2},
\cdots\biggr\}_{h=\frac{3}{2}}
\rightarrow\;\;\;\;\;\;\;\;
\biggl\{\frac{10}{21},\frac{32}{21},\frac{52}{21},\frac{74}{21},
\cdots\biggr\}_{h=\frac{32}{21}}
\rightarrow\nonumber\\
&&\biggl\{\frac{9}{19},\frac{29}{19},\frac{47}{19},\frac{67}{19},
\cdots\biggr\}_{h=\frac{29}{19}}
\rightarrow
\biggl\{\frac{8}{17},\frac{26}{17},\frac{42}{17},\frac{60}{17},
\cdots\biggr\}_{h=\frac{26}{17}}
\rightarrow
\biggl\{\frac{7}{15},\frac{23}{15},\frac{37}{15},\frac{53}{15},
\cdots\biggr\}_{h=\frac{23}{15}}
\rightarrow\nonumber\\
&&\biggl\{\frac{6}{13},\frac{20}{13},\frac{32}{13},\frac{46}{13},
\cdots\biggr\}_{h=\frac{20}{13}}
\rightarrow
\biggl\{\frac{5}{11},\frac{17}{11},\frac{27}{11},\frac{39}{11},
\cdots\biggr\}_{h=\frac{17}{11}}
\rightarrow
\biggl\{\frac{4}{9},\frac{14}{9},\frac{22}{9},\frac{32}{9},
\cdots\biggr\}_{h=\frac{14}{9}}
\rightarrow\nonumber\\
&&\biggl\{\frac{3}{7},\frac{11}{7},\frac{17}{7},\frac{25}{7},
\cdots\biggr\}_{h=\frac{11}{7}}
\rightarrow\;\;
\biggl\{\frac{2}{5},\frac{8}{5},\frac{12}{5},\frac{18}{5},
\cdots\biggr\}_{h=\frac{8}{5}}
\rightarrow\;\;\;\;\;
\biggl\{\frac{5}{13},\frac{21}{13},\frac{31}{13},\frac{47}{13},
\cdots\biggr\}_{h=\frac{21}{13}}
\rightarrow\nonumber\\
&&\biggl\{\frac{3}{8},\frac{13}{8},\frac{19}{8},\frac{29}{8},
\cdots\biggr\}_{h=\frac{13}{8}}
\rightarrow\;\;
\biggl\{\frac{4}{11},\frac{18}{11},\frac{26}{11},\frac{40}{11},
\cdots\biggr\}_{h=\frac{18}{11}}
\rightarrow
\biggl\{\frac{5}{14},\frac{23}{14},\frac{33}{14},\frac{51}{14},
\cdots\biggr\}_{h=\frac{23}{14}}
\rightarrow\nonumber\\
&&\biggl\{\frac{6}{17},\frac{28}{17},\frac{40}{17},\frac{62}{17},
\cdots\biggr\}_{h=\frac{28}{17}}
\rightarrow
\biggl\{\frac{1}{3},\frac{5}{3},\frac{7}{3},\frac{11}{3},
\cdots\biggr\}_{h=\frac{5}{3}}
\rightarrow\;\;\;\;\;\;\;
\biggl\{\frac{4}{13},\frac{22}{13},\frac{30}{13},\frac{48}{13},
\cdots\biggr\}_{h=\frac{22}{13}}
\rightarrow\nonumber\\
&&\biggl\{\frac{3}{10},\frac{17}{10},\frac{23}{10},\frac{37}{10},
\cdots\biggr\}_{h=\frac{17}{10}}
\rightarrow
\biggl\{\frac{5}{17},\frac{29}{17},\frac{39}{17},\frac{63}{17},
\cdots\biggr\}_{h=\frac{29}{17}}
\rightarrow
\biggl\{\frac{2}{7},\frac{12}{7},\frac{16}{7},\frac{26}{7},
\cdots\biggr\}_{h=\frac{12}{7}}\rightarrow\nonumber\\
&&\biggl\{\frac{3}{11},\frac{19}{11},\frac{25}{11},\frac{41}{11},
\cdots\biggr\}_{h=\frac{19}{11}}
\rightarrow
\biggl\{\frac{4}{15},\frac{26}{15},\frac{34}{15},\frac{56}{15},
\cdots\biggr\}_{h=\frac{26}{15}}
\rightarrow
\biggl\{\frac{5}{19},\frac{33}{19},\frac{43}{19},\frac{71}{19},
\cdots\biggr\}_{h=\frac{33}{19}}\rightarrow\nonumber\\
&&\biggl\{\frac{6}{23},\frac{40}{23},\frac{52}{23},\frac{86}{23},
\cdots\biggr\}_{h=\frac{40}{23}}
\rightarrow
\biggl\{\frac{1}{4},\frac{7}{4},\frac{9}{4},\frac{15}{4},
\cdots\biggr\}_{h=\frac{7}{4}}
\rightarrow\;\;\;\;\;\;\;
\biggl\{\frac{6}{25},\frac{44}{25},\frac{56}{25},\frac{94}{25},
\cdots\biggr\}_{h=\frac{44}{25}}\rightarrow\nonumber\\
&&\biggl\{\frac{5}{21},\frac{37}{21},\frac{47}{21},\frac{79}{21},
\cdots\biggr\}_{h=\frac{37}{21}}
\rightarrow
\biggl\{\frac{4}{17},\frac{30}{17},\frac{38}{17},\frac{64}{17},
\cdots\biggr\}_{h=\frac{30}{17}}
\rightarrow
\biggl\{\frac{3}{13},\frac{23}{13},\frac{29}{13},\frac{49}{13},
\cdots\biggr\}_{h=\frac{23}{13}}\rightarrow\nonumber\\
&&\biggl\{\frac{2}{9},\frac{16}{9},\frac{20}{9},\frac{34}{9},
\cdots\biggr\}_{h=\frac{16}{9}}
\rightarrow\;\;
\biggl\{\frac{1}{5},\frac{9}{5},\frac{11}{5},\frac{19}{5},
\cdots\biggr\}_{h=\frac{9}{5}}
\rightarrow\;\;\;\;\;
\biggl\{\frac{2}{11},\frac{20}{11},\frac{24}{11},\frac{42}{11},
\cdots\biggr\}_{h=\frac{20}{11}}\rightarrow\nonumber\\
&&\biggl\{\frac{3}{17},\frac{31}{17},\frac{37}{17},\frac{65}{17},
\cdots\biggr\}_{h=\frac{31}{17}}
\rightarrow
\biggl\{\frac{4}{23},\frac{42}{23},\frac{50}{23},\frac{88}{23},
\cdots\biggr\}_{h=\frac{42}{23}}
\rightarrow\;
\biggl\{\frac{1}{6},\frac{11}{6},\frac{13}{6},\frac{23}{6},
\cdots\biggr\}_{h=\frac{11}{6}}\rightarrow\nonumber\\
&&\biggl\{\frac{3}{19},\frac{35}{19},\frac{41}{19},\frac{73}{19},
\cdots\biggr\}_{h=\frac{35}{19}}
\rightarrow
\biggl\{\frac{2}{13},\frac{24}{13},\frac{28}{13},\frac{50}{13},
\cdots\biggr\}_{h=\frac{24}{13}}
\rightarrow\;
\biggl\{\frac{1}{7},\frac{13}{7},\frac{15}{7},\frac{27}{7},
\cdots\biggr\}_{h=\frac{13}{7}}\rightarrow\nonumber\\
&&\biggl\{\frac{2}{15},\frac{28}{15},\frac{32}{15},\frac{58}{15},
\cdots\biggr\}_{h=\frac{28}{15}}
\rightarrow
\biggl\{\frac{1}{8},\frac{15}{8},\frac{17}{8},\frac{31}{8},
\cdots\biggr\}_{h=\frac{15}{8}}
\rightarrow\;\;\;
\biggl\{\frac{2}{17},\frac{32}{17},\frac{36}{17},\frac{66}{17},
\cdots\biggr\}_{h=\frac{32}{17}}\rightarrow\nonumber\\
&&\biggl\{\frac{1}{9},\frac{17}{9},\frac{19}{9},\frac{35}{9},
\cdots\biggr\}_{h=\frac{17}{9}}.\nonumber
\end{eqnarray}

\end{document}